\documentclass[aps,twocolumn,nofootinbib,groupedaddress,superscriptaddress,longbibliography,notitlepage]{revtex4-2}

\usepackage{amsmath,amssymb}
\usepackage[normalem]{ulem}
\usepackage[inkscapelatex=false]{svg}
\usepackage{graphicx}
\usepackage{longtable,multirow}
\usepackage{mathtools}
\usepackage{dsfont}
\usepackage{amsfonts}
\usepackage{xcolor}
\usepackage{url}
\usepackage{braket}
\usepackage{array}
\usepackage{booktabs}
\usepackage[colorlinks=true,linkcolor=teal,citecolor=teal,urlcolor=teal,hypertexnames=false]{hyperref}
\newcommand{\nocontentsline}[3]{}
\newcommand{\tocless}[2]{\bgroup\let\addcontentsline=\nocontentsline#1{#2}\egroup}
\def\ba#1\ea{\begin{align}#1\end{align}}
\def\bg#1\eg{\begin{gather}#1\end{gather}}
\def\bpm{\begin{pmatrix}}
\def\epm{\end{pmatrix}}
\def\bbm{\begin{bmatrix}}
\def\ebm{\end{bmatrix}}

\newcommand{\teal}[1]{{\color{teal} #1}}

\allowdisplaybreaks
\begin{document}
\title{Anomalous suppression of quantum chaos between two integrable limits}
\author{Roopayan Ghosh}
\thanks{These authors contributed equally to this work.}
\affiliation{School of Basic Sciences, Indian Institute of Technology, Bhubaneswar, 752050, India}
\affiliation{Department of Physics and Astronomy, University College London, Gower Street, WC1E 6BT, London, United Kingdom}
\author{Henry Davenport}
\thanks{These authors contributed equally to this work.}
\affiliation{Blackett Laboratory, Imperial College London, London SW7 2AZ, United Kingdom}
\author{Frank Schindler}
\affiliation{Blackett Laboratory, Imperial College London, London SW7 2AZ, United Kingdom}

\begin{abstract}
Level statistics in non-integrable quantum many-body systems with time reversal symmetry are expected to follow the Gaussian Orthogonal Ensemble (GOE), a hallmark of quantum chaos. However, we show that the interacting Su-Schrieffer-Heeger model exhibits a clear suppression of the mean level-spacing ratio $\langle r\rangle$ from the GOE value $\approx 0.535$, persisting deep in the nonintegrable regime. This challenges the conventional association between non-integrability and fully chaotic spectral statistics. Using exact diagonalization supported by semi-analytical arguments, we trace this anomaly to incomplete hybridization of many-body band states inherited from the noninteracting band structure. The resulting restructuring of the spectrum weakens level repulsion without restoring integrability. We show the robustness of this mechanism in extensions of the model which break chiral and inversion symmetry.
\end{abstract}

\let\oldaddcontentsline\addcontentsline
\renewcommand{\addcontentsline}[3]{}
\maketitle

\teal{\textit{Introduction---}} Understanding the emergence of ergodicity in isolated quantum systems is a central problem in many-body physics. In generic interacting systems, the eigenstate thermalization hypothesis (ETH) predicts that generic eigenstates behave thermally, leading to universal spectral correlations described by random-matrix theory~\cite{Deutsch1991,Srednicki1994,Bogmolny}, which is a hallmark of quantum chaos. Deviations from this behavior, often signaled by Poisson level statistics, indicate a breakdown of ergodicity and are typically associated with mechanisms such as the putative disorder-induced many-body localization~\cite{Evers2008Anderson,Basko06,gornyi2005interacting,Abanin_RMP,PhysRevB.77.064426}, quantum many-body scars~\cite{Turner2018,Moudgalya_2022}, or Hilbert-space fragmentation due to kinetic constraints~\cite{PhysRevX.10.011047}.

These phenomena share a common feature: the suppression of ergodicity is usually tied to identifiable structures such as disorder, conserved quantities, or constrained dynamics. It is therefore natural to ask whether nonergodic spectral signatures can arise in clean, translationally invariant systems without such ingredients, and if so, what mechanism controls their emergence.

In this work, we demonstrate that adding interactions even in simple two-band tight binding models can cause deviations from random-matrix behavior deep within a non-integrable regime. As an example, we study the interacting Su-Schrieffer-Heeger (SSH) chain~\cite{PhysRevB.110.165145,PhysRevB.108.195151,Chang2025} under periodic boundary conditions (PBC), where the noninteracting model is translationally invariant with two-bands. By analyzing the full many-body (MB) eigenspectrum, we uncover a pronounced line of suppressed level repulsion embedded within an otherwise ergodic phase.

We show that this anomaly does not arise from integrability or localization, but from incomplete hybridization between states originating from different MB bands. Using perturbative arguments, we demonstrate that the underlying single-particle band structure strongly suppresses hybridization near the integrable limits, and remarkably this effect persists deep into the non-integrable regime. Our results therefore reveal a distinct mechanism for suppressing quantum chaos in clean systems, driven by band structure rather than disorder or conventional dynamical constraints.
\paragraph*{\teal{Model---}}
A simple example of a system which exhibits anomalous suppression of quantum chaos is the interacting 1D spinless SSH model. The non-interacting SSH Hamiltonian is defined in terms of creation/annihilation operators $c^\dagger_{R, i}/c_{R, i}$ for spinless fermions at unit cell $R$, orbital $i\in \{A, B\}$,~\cite{PhysRevLett.42.1698}
\begin{equation}
\hat{H}_0 = \sum_R (1 + \delta)c_{R, A}^{\dagger} c_{R, B} + (1 - \delta)c_{R, B}^{\dagger} c_{R+1, A} + \mathrm{H.c}.
\label{eq:SSH}
\end{equation}
The real parameter $\delta$ controls the alternating hopping. We study the model exclusively in periodic boundary conditions~(PBCs). Since this has two sites per unit cell, Fourier transforming Eq.~\eqref{eq:SSH} yields a two-band model with the single-particle dispersion,
\begin{equation}
E_{\pm}(k)=\pm\sqrt{2\left[\left(1+\delta^2\right)+(1-\delta^2)\cos k\right]}.\label{eq:singleparticledispersion}
\end{equation}
We can write the initial Hamiltonian in the band basis as $\hat{H}_0  = \sum_{k, \alpha\in \{+, -\}} E_{\alpha}(k) c^\dagger_{k, \alpha} c_{k, \alpha}$, where $ c^\dagger_{k, \alpha} /c_{k, \alpha}$ is the creation/annihilation operator for a fermion in band $\alpha \in \{+, -\}$. 
To break the integrability of the model, we study 
\begin{align}
\hat{H} &= \hat{H}_0 + \hat{H}_{\mathrm{int}}\label{eq:genSSH}, \\ \hat{H}_{\mathrm{int}} &= V_{\mathrm{int}}\sum_R (n_{R, A} n_{R, B}+ n_{R, B}n_{R+1, A}),
\end{align}
where we have added nearest-neighbor density-density interactions, with $n_{R, i}=c_{R, i}^\dagger c_{R, i}$. These are the simplest interactions for a spinless model. This modification yields a nontrivial correlated problem whose global spectral properties remain largely unexplored.
\begin{figure}

    \centering
   \includegraphics[width=\columnwidth]{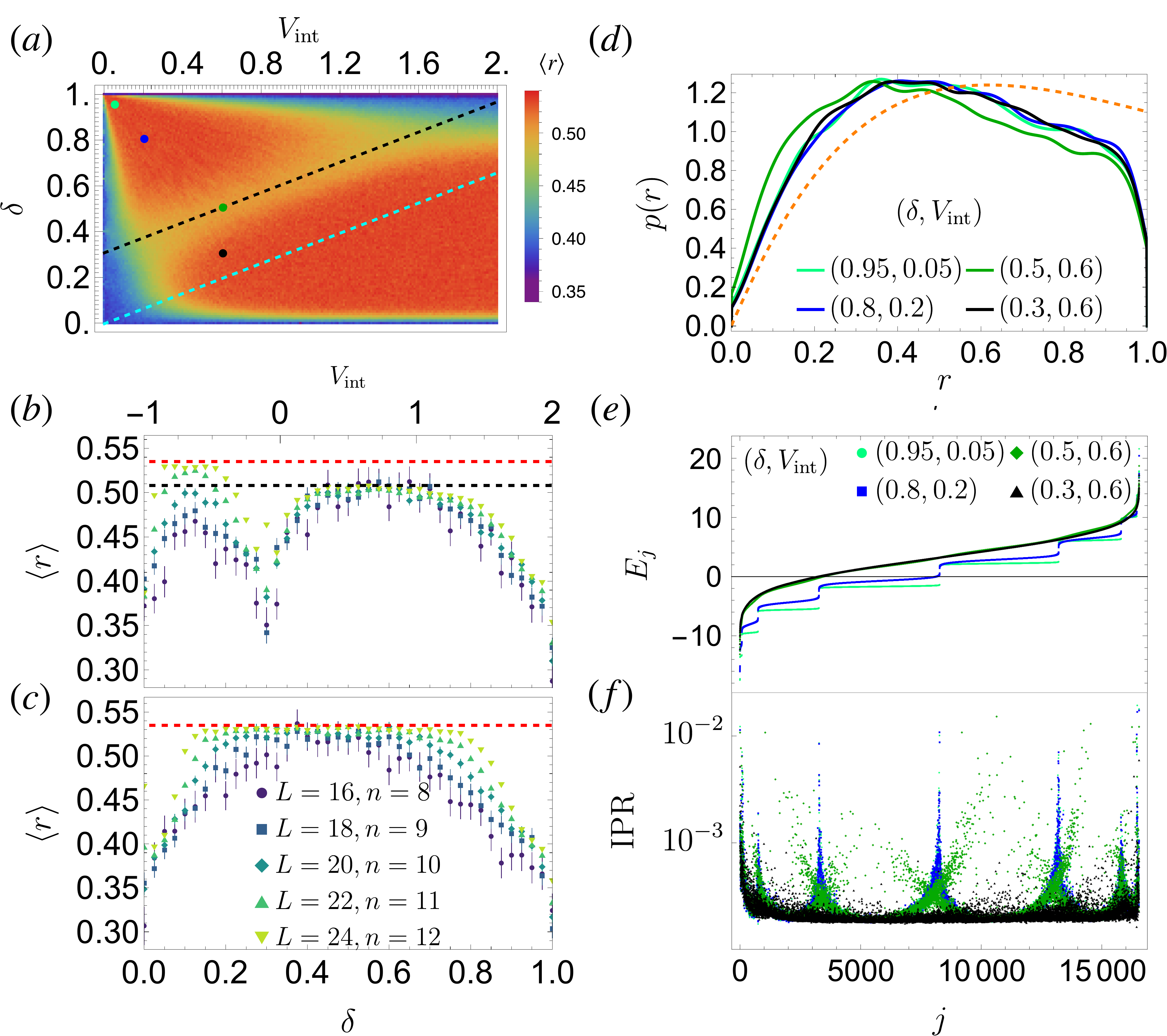}  \\

   \caption{(a)  Adjacent gap ratio $\langle r \rangle$ for the model in Eq.~\eqref{eq:genSSH} for a $L=22$, half filled chain with $n=11$ electrons over a broad parameter regime, exhibiting a pronounced anomalous dip. The black dashed line indicates an approximate linear fit through this anomalous region, while the cyan dashed line is drawn parallel to it through a representative generic region. (b) $\langle r \rangle$ along the cut $\delta = 1/3 + V_{\mathrm{int}}/3$ (black dashed line in the left panel). (c) $\langle r \rangle$ along the cut $\delta = V_{\mathrm{int}}/3$ (cyan dashed line). Different colors denote system sizes $L=16$–$24$ at half filling. (d) Probability distribution of consecutive level spacing ratio $r$ comparing four representative points from Fig.~\ref{fig:pbcrstat}(a) for $L=24$ at half filling. The red dashed line denotes typical  distribution for GOE statistics.(e) Comparison of the eigenspectrum of the same points, $j$ labels an eigenstate. (f) Distribution of IPR (Eq.~\eqref{eq:IPR}) of eigenstates labeled by $j$ for the same parameters. 
   }
    \label{fig:pbcrstat}
\end{figure}

\textit{\teal{Spectral signatures---}} To characterize the full eigenspectrum of $H$, we first employ level statistics~\cite{PalHuse,PhysRevB.99.104205} as a probe of ergodicity and its variation across the MB spectrum, which has also been recently experimentally accessed in a spin model~\cite{vandersypenlevelstat}. Our primary diagnostic is the adjacent level-spacing ratio~\cite{Atas2013distribution, Evers2008Anderson}, supplemented by eigenstate-resolved measures such as the inverse participation ratio (IPR). Given the ordered energies $\{\epsilon_i\}$, we define
\begin{equation}
\delta_i = \epsilon_{i+1} - \epsilon_i, \qquad 
r_i = \frac{\min(\delta_i, \delta_{i+1})}{\max(\delta_i, \delta_{i+1})},
\label{eq:rstat}
\end{equation}
where $r_i$ is called the level spacing ratio. Chaotic many-body systems with time-reversal symmetry exhibit Gaussian Orthogonal Ensemble (GOE) statistics with $\langle r\rangle\approx0.535$, while integrable or localized spectra exhibit Poisson statistics with $\langle r\rangle\approx0.386$. Eq.~\eqref{eq:genSSH} preserves time-reversal symmetry represented by the \emph{anti-unitary} operator $\hat{\mathcal{T}}$ that acts on the fermion creation operators as $\hat{\mathcal{T}} c^\dagger_{R, A/B}\hat{\mathcal{T}}^{-1} = c^\dagger_{R, A/B}$ with $\hat{\mathcal{T}}^2 = 1$. Therefore, GOE provides the appropriate random-matrix benchmark throughout this work. These values are meaningful only after resolving all microscopic unitary symmetries. Intermediate values of $\langle r\rangle$ can indicate either unresolved symmetries~\cite{PhysRevX.12.011006}, or genuine non-ergodicity~\cite{PhysRevE.77.035201,PhysRevB.107.115424}.

For the interacting SSH chain with PBC, we compute level statistics within a fully symmetry-resolved sector, fixing momentum ($p=0$), parity (even), and also resolving the particle-hole symmetry at half-filling to eliminate artificial reductions of $\langle r\rangle$. Figure~\ref{fig:pbcrstat}(a) shows the resulting mean level-spacing ratio for Eq.~\eqref{eq:genSSH} at $L=22$.

Fig.~\ref{fig:pbcrstat}(a) shows several expected structures: along the integrable lines $V_{\rm int}=0$, $\delta=0$, and $\delta=1$, we find $\langle r\rangle \approx 0.38$, consistent with Poisson statistics. The system is integrable for $V_{\rm int}=0$ since it reduces to the non-interacting SSH model. The model with constant hopping ($\delta=0$) is integrable since it can be mapped to the Heisenberg chain via a Jordan-Wigner transformation~\cite{LIEB1961407, JordanWignerTransformation}. Showing integrability at $\delta=1$ is more subtle. In this fully dimerized limit, every other hopping vanishes, producing an extensive set of conserved dimer occupations and hence Poisson statistics. The model is again exactly solvable through a mapping to the transverse-field Ising chain (Appendix~\ref{sec:TFIM})~\cite{10.21468/SciPostPhysCore.6.4.083}. At larger interactions (not shown), $\langle r\rangle$ gradually decreases toward Poisson values, consistent with interaction-induced approximate Hilbert-space fragmentation; details are discussed in Appendix~\ref{app:largeV}.

Away from these limits, the spectrum is predominantly ergodic. However, we observe a robust line of suppressed $\langle r\rangle$ deep within the non-integrable regime. A linear fit gives the empirical relation $\delta = 0.3 + 0.366\,V_{\rm int}\approx \frac{1}{3}(1+V_{\rm int})$, shown as the black dashed line in Fig.~\ref{fig:pbcrstat}(a).

We now examine the finite-size dependence of this feature by computing $\langle r\rangle$ along the anomalous line for several system sizes. For comparison, we also consider a reference line that avoids the anomalous region [cyan dashed line in Fig.~\ref{fig:pbcrstat}(a)]. Along the reference line [Fig.~\ref{fig:pbcrstat}(c)], $\langle r\rangle$ approaches the GOE value $\approx0.535$ (red dashed line) away from the integrable limits. In contrast, along the anomalous line [Fig.~\ref{fig:pbcrstat}(b)], $\langle r\rangle$ remains systematically reduced for $1/3<\delta<1$, saturating near $\langle r\rangle\approx0.508$ (black dashed line) up to the largest accessible system size, $L=24$ (Hilbert-space dimension $\mathcal{D}=57704$). We therefore find no significant finite-size drift of the anomaly within the accessible range, suggesting that the observed suppression is not a simple finite-size effect but originates from an intrinsic structural feature of the interacting spectrum.

To identify the origin of the anomalous $\langle r\rangle$, we examine the evolution of the many-body spectrum and inverse participation ratio (IPR) across the parameter sequence 
$(\delta,V_{\rm int})=(0.95,0.05)\rightarrow(0.8,0.2)\rightarrow(0.5,0.6)\rightarrow(0.3,0.6)$, 
which interpolates between the nearly dimerized flat-band limit and the ergodic regime while crossing the anomalous line [see the corresponding points marked on Fig.~\ref{fig:pbcrstat}(a)]. Figures~\ref{fig:pbcrstat}(d)-(f) show the corresponding evolution of level statistics, eigenenergy structure, and eigenstate properties.

We first examine the full distribution $p(r)$ of adjacent level-spacing ratios [Fig.~\ref{fig:pbcrstat}(d)]. The ergodic points follow GOE statistics up to finite size corrections with $p(r=0)\approx0$, including the point closest to the integrable dimerized limit. In contrast, the point $(0.5,0.6)$ on the anomalous line exhibits a pronounced enhancement near $r\simeq0$, indicating weakened level repulsion despite being far from any integrable limit.

To understand the origin of the weakened level repulsion, we examine the energy eigenspectrum [Fig.~\ref{fig:pbcrstat}(e)]. In the non-interacting dimerized limit ($\delta\sim1$, $V_{\rm int}\sim0$), the single-particle spectrum consists of two flat bands separated by a gap, producing exactly flat MB bands. The light green points in Fig.~\ref{fig:pbcrstat}(e), corresponding to $\delta=0.95$ and $V_{\rm int}=0.05$, show that weak hopping and interactions broaden these bands slightly while keeping them spectrally separated at $L=24$. Increasing interactions and/or reducing dimerization further broadens the MB bands. At $\delta=0.8$ and $V_{\rm int}=0.2$, the MB band widths increase substantially and several gaps nearly close. At the representative point $\delta=0.5$ and $V_{\rm int}=0.6$ on the anomalous line [green points in Fig.~\ref{fig:pbcrstat}(e)], the MB band gaps have closed completely. Finally, we look at a parameter choice ($\delta = 0.3, V_{\mathrm{int}} = 0.6$) on the other side of the anomalous line to the initial two points, where we see a continuous spectrum as well. We therefore conclude that the anomalous regime emerges after the MB band gaps close, when states originating from different MB bands begin to overlap. We next use the IPR to probe the degree of hybridization between these states.

The IPR is defined as
\begin{equation}
\mathrm{IPR}_j = \sum_{k=1}^{\mathcal{D}} 
|\langle k | E_j \rangle|^4,
\label{eq:IPR}
\end{equation}
where $\ket{E_j}$ is an eigenstate and $\{\ket{k}\}$ denotes the computational basis of the real-space occupation-number configurations. Ergodic states satisfy $\mathrm{IPR}\sim\mathcal{O}(1/\mathcal{D})$, where $\mathcal{D}$ is the Hilbert space dimension, whereas states with reduced ergodicity yield significantly larger values.

Along the progression described above, the nearly flat-band limit displays well-separated MB bands and enhanced IPR near their edges. Since adjacent MB bands remain spectrally isolated in this regime, nearby eigenstates predominantly originate from the same MB band of the dimerised limit and can hybridize efficiently within it, producing GOE-like level statistics. At intermediate parameters, the MB bands broaden and begin to overlap, while remnants of the original band structure remain visible in both the spectrum and IPR.
In fact, comparing Fig.~\ref{fig:pbcrstat}(e) and (f) we can see that at $(\delta=0.5,V_{\rm int}=0.6)$, the visible MB band gaps have closed, yet enhanced IPR persists near the merging points of adjacent MB bands. These states are remnants of the states at the edges of the MB bands in the dimerized limit, which were gapped. Despite the fact that they become degenerate with states originating from the other MB band, they remain weakly hybridized with them. We conclude that, before the MB bands overlap, hybridization within individual MB bands produces ordinary level repulsion. However, after the MB bands overlap, near-degenerate states that originate from different MB bands become abundant but remain only weakly hybridized, suppressing $\langle r\rangle$ without inducing localization.

Finally, at $(0.3,0.6)$, MB band remnants disappear entirely, $\mathrm{IPR}\approx 3/\mathcal{D}$, and $\langle r\rangle$ returns to its GOE value $\approx 0.535$. The persistence of weakly hybridized states in the anomalous region however points to the importance of analyzing the structure of eigenstates in the integrable limits to find what restricts their hybridization.

\textit{\teal{Perturbative mechanism for anomalous $\langle r\rangle$ statistics---}} We now develop a perturbative understanding of the level statistics anomaly near the two integrable endpoints of the anomalous line: the non-interacting limit ($V_{\rm int}=0$) and the fully dimerized limit ($\delta=1$). We begin with the latter case. Here, the states at the top and bottom of adjacent many-body minibands have qualitatively different structures, so hybridization between them occurs only at high order in perturbation theory, strongly suppressing level repulsion.

In the dimerized limit with $V_{\mathrm{int}} = 0$, the Hamiltonian decomposes into isolated dimers, each with bonding and antibonding single-particle orbitals of energies $\mp2$. At fixed filling, we obtain flat MB bands consisting of states with the same number of occupied antibonding orbitals, $n_+$. This simultaneously sets the number of bonding orbitals ($n_-$) since $n_++n_- = n$ \emph{i.e.} the total number of electrons. Each flat MB band contains states that have the same energy but differ in their real-space structure — specifically, in how the bonding and antibonding orbitals are distributed amongst the different dimers. For example, states with doubly occupied and empty dimer pairs are degenerate with states in which an electron has been excited from a bonding to an antibonding orbital within a single dimer — both configurations cost the same energy in the non-interacting limit.

When interactions are added, the MB bands acquire a finite width. However, the dimerised ($\delta=1$) limit still retains an extensive set of conserved dimer occupations, allowing the states at the top and bottom of each MB band to be characterized exactly. States at the top maximize the interaction energy and therefore contain as many (neighboring) doubly occupied dimers as possible (Appendix~\ref{app:doubleOccDimersProof}). Since $n_+$ and $n_-$ are fixed within a given MB band, the maximum number of doubly occupied dimers is $\min(n_+,n_-)$. In contrast, states at the bottom minimize the interaction energy and therefore contain one electron per dimer.

As $V_{\rm int}$ increases along the $\delta=1$ axis, the MB bands broaden and eventually begin to overlap. Consider two adjacent MB bands at the onset of overlap. The state at the bottom of the upper MB band contains $\min(n_+,n_-)$ doubly occupied dimers, while the state at the top of the lower MB band has one electron per dimer. Because these states have different conserved dimer occupations, the interaction alone cannot couple them. Hybridization therefore requires inter-dimer hopping, which we treat perturbatively by setting $\delta=1-\epsilon$. To connect the two states, at least $\ell\ge\min(n_+,n_-)$ electrons must hop out of doubly occupied dimers, so the effective level splitting first appears at $\ell$-th order in perturbation theory (also verified numerically),
\begin{equation}
t_{\rm eff}\sim \frac{\epsilon^\ell}{V_{\rm int}^{\ell-1}}.
\label{eq:dimerizedMatrixElement}
\end{equation}
Here, the numerator arises from inter-dimer hopping processes, while the denominators correspond to the energies of the intermediate virtual states. Since each hop reduces the number of doubly occupied dimers by one, the intermediate-state energies differ by $\sim V_{\rm int}$ (Appendix~\ref{app:doubleOccDimersProof}). The resulting effective coupling is therefore strongly suppressed, leading to weak level repulsion between overlapping MB bands. Although we have focused on the extremal states of adjacent MB bands, the same mechanism applies to a large class of states near the band edges. As the MB bands broaden further, however, states connected at lower perturbative order become energetically accessible, leading to the eventual recovery of GOE statistics as $\delta$ decreases away from $1$.

We now turn to the second integrable endpoint of the anomalous line, the non-interacting limit $V_{\rm int}=0$. Here, the ground state consists of all electrons occupying the lower single-particle band, while the $n^{\rm th}$ MB band contains $n$ electron-hole excitations across the band gap. At $\delta=1$ these MB bands are flat and gapped, but decreasing $\delta$ broadens them and eventually causes them to overlap. We therefore ask whether the resulting overlapping states hybridize once perturbative interactions are introduced. 

To address this, we first characterize the states at the edges of adjacent MB bands. Since the single-particle gap is minimal at $k=\pi$ and maximal at $k=0$ [Eq.~\eqref{eq:singleparticledispersion}], states at the top of the $n^{\rm th}$ MB band contain $n$ electron-hole excitations localized near $k=0$, whereas states at the bottom of the $(n+1)^{\rm th}$ band contain $n+1$ excitations near $k=\pi$. When weak interactions of strength $V_{\rm int}$ are added, one might expect the overlapping MB bands to immediately hybridize. However, this hybridization remains strongly suppressed. Rewriting the interaction in the band basis,
\begin{eqnarray}
\hat H_{\rm int}
=
\sum_{k,q,p}
\sum_{\alpha,\beta,\gamma,\zeta=\pm}
f^{k,q,p}_{\alpha,\beta,\gamma,\zeta}
c^\dagger_{k,\alpha}
c_{q,\beta}
c^\dagger_{p,\gamma}
c_{k-q+p,\zeta},
\end{eqnarray}
where $\alpha,\beta,\gamma,\zeta$ label the two bands and $f^{k,q,p}_{\alpha,\beta,\gamma,\zeta}$ is a form factor. A single application of $\hat H_{\rm int}$ can scatter at most one electron-hole pair from $k=0$ to $k=\pi$, \textit{e.g.}
$c^\dagger_{\pi,+}c_{\pi,-}c^\dagger_{0,-}c_{0,+}$. Consequently, hybridizing a state with $n$ electron-hole excitations near $k=0$ with one containing $n+1$ excitations near $k=\pi$ requires at least order-$n$ perturbation theory (strictly, order $n+1$ after accounting for the additional particle-hole excitation). The resulting effective matrix element therefore scales (also checked numerically) as
\begin{eqnarray}
t_{\rm eff}
\sim
\frac{(V_{\rm int})^n}{(1-\delta)^{n-1}},
\end{eqnarray}
where the denominators arise from the intermediate-state energy gaps. These gaps are set by the energy difference between electron-hole excitations at $k=0$ and $k=\pi$ (Appendix~\ref{app:pert}), which is of order the single-particle bandwidth $\sim(1-\delta)$. Since the MB bands first overlap while this bandwidth remains large, hybridization between these states remains parametrically suppressed.

\begin{figure}[t]
\centering 

\includegraphics[width=\columnwidth]{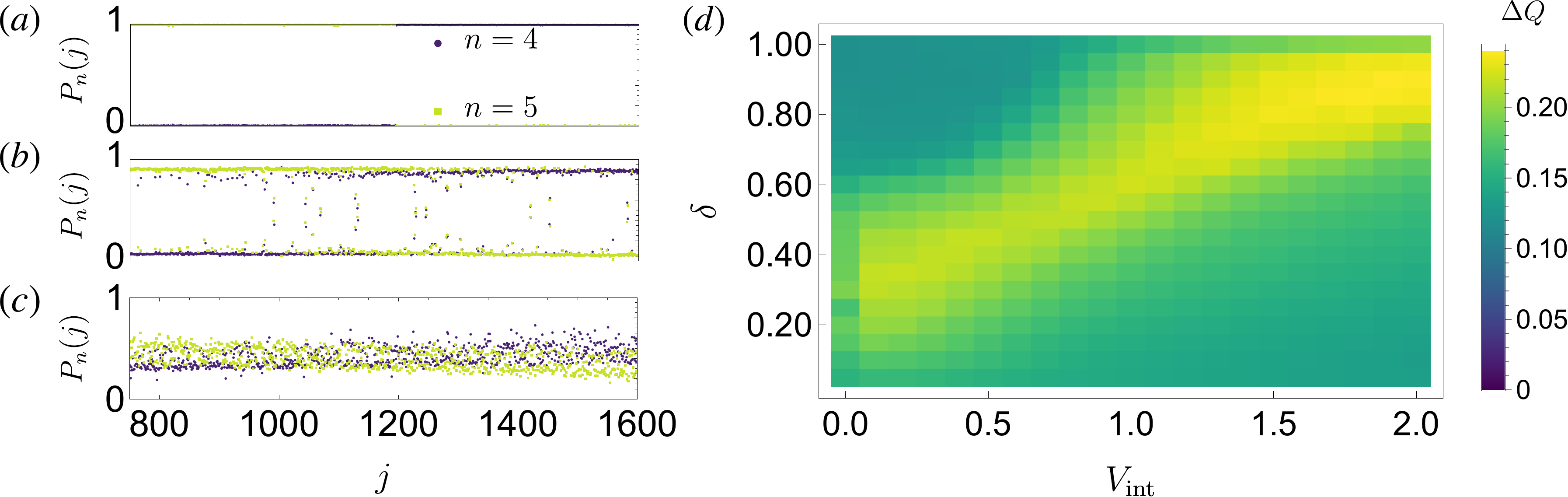}
\caption{$(a), (b), (c)$ Expectation of the $j^{\mathrm{th}}$ eigenstate in the projector onto the $n^{\mathrm{th}}$ dimerised limit MB band (denoted $P_{n}(j)$) for $n=4,5$. Eigenstates calculated for $L = 22$, $V = 0.6$ and $(a)$ $\delta = 0.95$, $(b)$ $\delta = 0.45$ and $(c)$ $\delta = 0.05$. $(d)$ Standard deviation $\Delta Q$ of $Q=Q(j)$ defined by Eq.~\eqref{eq:Diagnostic} showing a rise approximately at regions of anomalously low $\langle r \rangle$.
} 
\label{fig:variance} 
\end{figure}
\textit{\teal{Weak hybridization beyond the perturbative regime---}} We now show numerically that incomplete hybridization between adjacent MB bands persists even far from the integrable limits. To quantify this, we project the exact eigenstates onto the MB bands of the dimerized limit. Denoting the projector onto the $n^{\rm th}$ MB band by $\hat P_n$, the overlap of the eigenstate $|E_j\rangle$ with this band is
\begin{equation}
P_n(j)\equiv \langle E_j|\hat P_n|E_j\rangle.
\end{equation}

In the dimerized limit, $P_n(j)$ is exactly $0$ or $1$ depending on the band character of the state [Fig.~\ref{fig:variance}(a)], whereas complete hybridization between adjacent MB bands would yield a nearly uniform intermediate value $\sim1/2$ [Fig.~\ref{fig:variance}(c)]. Instead, in the anomalous regime we observe strong fluctuations in band character between adjacent eigenstates [Fig.~\ref{fig:variance}(b)], indicating that states with similar energies remain only weakly hybridized.

To quantify this effect, we define
\begin{equation}
Q(j)=P_n(j)P_{n+1}(j+1),
\label{eq:Diagnostic}
\end{equation}
where $n$ and $n+1$ denote adjacent MB bands in the dimerized limit. This quantity is close to $0$ when consecutive eigenstates originate predominantly from the same MB band, close to $1$ when they originate from different MB bands, and takes intermediate values once strong hybridization occurs. Consequently, $Q(j)$ is nearly constant both when adjacent MB bands remain fully separated [Fig.~\ref{fig:variance}(a)] and when they are fully hybridized [Fig.~\ref{fig:variance}(c)]. In contrast, coexistence of weakly hybridized states from different MB bands produces strong fluctuations of $Q(j)$ between $0$ and $1$.

We characterize these fluctuations using the standard deviation $\Delta Q=\sqrt{\mathrm{Var}[Q(j)]}$. As shown in Fig.~\ref{fig:variance}(d), $\Delta Q$ exhibits a pronounced maximum along the anomalous line where $\langle r\rangle$ is minimized, demonstrating that the suppression of level repulsion originates from incomplete hybridization between different MB bands.

\textit{\teal{Discussion---}} We have identified a robust line of suppressed level repulsion, characterized by reduced $\langle r\rangle$, embedded deep within the interacting and non-integrable regime of the SSH chain. Using analytic arguments and exact diagonalization, we showed that this suppression originates from incomplete hybridization between many-body states inherited from different minibands of the non-interacting spectrum. This reveals a distinct mechanism for suppressing quantum chaos in clean systems, fundamentally different from conventional scenarios based on integrability, disorder, or localization. In Appendix~\ref{app:SymmetryBreakingVariant}, we further show that this feature survives even after explicitly breaking the inversion and chiral symmetries of the SSH model, demonstrating that it is not symmetry protected.

We find that the anomaly persists up to the largest system sizes accessible to exact diagonalization. However, its fate in the thermodynamic limit remains unclear. As $L\rightarrow\infty$, the finite width of the single-particle bands causes the many-body minibands to overlap extensively, producing a continuum of states in the middle of the spectrum even in the non-interacting limit. Since the suppression originates from weakly hybridized remnants of these minibands, it remains an open question whether this regime survives in the thermodynamic limit or is eventually washed out by the many-body continuum.

Our results therefore identify a crossover regime between band-structured and fully ergodic spectra, where near-degenerate but weakly hybridized states suppress level repulsion without producing global localization. These atypical states may also leave observable dynamical signatures, including anomalously slow relaxation or long-lived prethermal behavior for initial states with strong overlap on the miniband remnants. More broadly, our work raises the question of whether similar intermediate-statistics regimes emerge generically in interacting multiband systems, and how band-geometry-induced constraints can be incorporated into generalized theories of quantum chaos and ergodicity.

 \paragraph*{Acknowledgements:} RG would like to thank Marko Znidaric, Sougato Bose and Madhumita Sarkar for fruitful discussions. HD would like to thank Johannes Knolle for useful discussions regarding the transverse field Ising model mapping. HD acknowledges support from the Engineering and Physical Sciences Research Council (grant number EP/W524323/1). This work was supported by a UKRI Future Leaders Fellowship MR/Y017331/1. 

%%%%%%Ref
\bibliography{ref}
%%%%%%

% %%%%%%Appendix
\let\addcontentsline\oldaddcontentsline

\clearpage
% %%%%%%

%%%%%%
\onecolumngrid
\begin{center}
\textbf{\large Supplemental Material for \\
``Anomalous suppression of quantum chaos in a multi-band model''}
\end{center}
%%%%%%

%%%%%%
\setcounter{section}{0}
\setcounter{figure}{0}
\setcounter{equation}{0}
\renewcommand{\thefigure}{S\arabic{figure}}
\renewcommand{\theequation}{S\arabic{equation}}
\renewcommand{\thesection}{S\arabic{section}}
\tableofcontents
%%%%%%

%%%%%%
\hfill \\
\newpage

 % \appendix
\section{Large $V_{\rm int}$ results for $\langle r \rangle$}
 \label{app:largeV}
 \begin{figure}[h]
    \centering
    \includegraphics[width=0.5 \columnwidth]{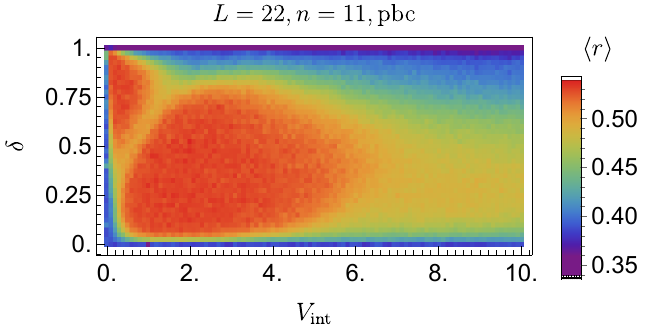}
    \caption{ $\langle r \rangle$ statistics of the model in Eq.~\eqref{eq:genSSH} in different parameter regimes}
    \label{fig:pbcrstatapp}
\end{figure}

In this section we present the behavior of $\langle r\rangle$ over the full parameter regime, including large $V_{\rm int}$ (see Fig.~\ref{fig:pbcrstatapp}). For sufficiently large interactions, $\langle r\rangle$ decreases toward Poisson values, indicating a breakdown of ergodicity. This behavior can be understood in terms of (approximate) Hilbert-space fragmentation~\cite{Moudgalya_2022}, similar to that discussed in Ref.~\cite{PhysRevB.109.045145}.

The fragmentation is approximate because different regions of Hilbert space remain weakly connected via high-order virtual processes. However, the associated tunneling amplitudes are parametrically suppressed in $V_{\rm int}$, effectively generating large energy barriers and strongly reducing hybridization between sectors.

A first level of fragmentation arises from the number of adjacent occupied pairs, denoted by $n_d$. For example, for $L=12$ and $n=6$, the Hilbert space decomposes into sectors labeled by $n_d$ with energies $\sim n_d V_{\rm int}$, with representative states, 
\begin{equation}
\begin{split}
&n_d=0:\quad \ket{101010101010}, \\
&n_d=1:\quad \ket{011010101010}, \\
&n_d=2:\quad \ket{011100101010}, \\
&n_d=3:\quad \ket{000111101010},\ \ket{000101110110}, \\
&n_d=4:\quad \ket{000111110010}, \\
&n_d=5:\quad \ket{000111111000}.
\end{split}
\end{equation}
This classification is closely related to the different dimer sectors discussed in the main text. In the strong-coupling limit, the number of adjacent occupied pairs $n_d$ directly controls the interaction energy, $E \sim n_d V_{\rm int}$, leading to a reorganization of the spectrum into well-separated sectors. While the miniband structure discussed at intermediate coupling is washed out by hybridization, large $V_{\rm int}$ induces an emergent energetic hierarchy between these sectors, effectively restoring a gap structure controlled by interactions.
States within the same $n_d=3$ sector may hybridize efficiently, while transitions between sectors are suppressed by an energy cost $\sim V_{\rm int}$. For example, the process
\begin{equation}
\ket{000111101010} \rightarrow \ket{000111101001}
\end{equation}
remains within the same $n_d$ sector and therefore hybridizes strongly. In contrast,
\begin{equation}
\ket{000111101010} \nrightarrow \ket{000111101100}
\end{equation}
changes $n_d$ and is suppressed by an energy cost $\sim V_{\rm int}$, leading to an effective hybridization $\sim (1\pm\delta)/V_{\rm int}$.

Even within a fixed $n_d$ sector, further fragmentation arises because different configurations are connected only via higher-order processes. For instance,
\begin{equation}
\ket{000111101010} \nrightarrow \ket{001011110010}
\end{equation}
requires intermediate states with $n_d=2$ or $n_d=4$, leading to second-order hybridization amplitudes $\sim (1\pm\delta)^2/V_{\rm int}^2$. Consequently, the connectivity of the Hilbert space becomes highly inhomogeneous.

This structure can be visualized by grouping local configurations into three types: 
$h=\ket{00}$, $d=\ket{11}$, and $p=\ket{01},\ket{10}$. A typical configuration can then be written schematically as
\begin{equation}
\ket{hhpdddppppph}.
\end{equation}
In this representation, $h$ and $d$ domains cannot pass through each other, while $p$ excitations move freely. This kinematic constraint partitions the Hilbert space into dynamically disconnected or weakly connected regions, giving rise to fragmentation.

As a consequence, eigenstates have highly non-uniform support in the computational basis, leading to enhanced inverse participation ratio (IPR), reduced entanglement entropy, and suppressed level repulsion. The resulting behavior is consistent with the reduction of $\langle r\rangle$ observed at large $V_{\rm int}$.
\newpage\section{Mapping to transverse field Ising model}\label{sec:TFIM}
\subsection{Model}
We consider an interacting generalisation of the SSH model. We study the SSH model in the dimerised limit (where we retain only \emph{intra}cell hopping not \emph{inter}cell). In addition we add Hubbard interactions between nearest neighbour sites, these are the simplest interactions since we are considering a spinless model. Consider a lattice with 2 atomic orbitals per unit cell labelled $A, B$. The annihilation operator $c_{R, i}$ annihilates an electron in unit cell $R$, site $i\in \{A, B\}$. We study the system in periodic boundary conditions with $L_{c}$ unit cells. The full Hamiltonian consists of a kinetic part $\hat{H}_0$ and interactions $\hat H_{\mathrm{int}}$,
\begin{align}
\hat H = \hat{H}_0 + \hat H_{\mathrm{int}}\label{eq:totalHamiltonian}
\end{align}
where,
\begin{align}
\hat{H}_0 = v \sum_{R} \left(c^\dagger_{R, A} c^{\vphantom{\dagger}}_{R, B} + c^\dagger_{R, B} c^{\vphantom{\dagger}}_{R, A}\right).
\end{align} 
For simplicity of notation we have defined $v = 1+\delta$. Similarly
\begin{align}
\hat{H}_{\mathrm{int}} =V_{\mathrm{int}}\sum_{R} \hat n_{R, A} \hat n_{R, B} + V_{\mathrm{int}} \sum_{R} \hat n_{R, B} \hat n_{R+1, A},
\end{align}
where $n_{R, i} = c^\dagger_{R, i} c_{R, i}$ is the number operator for unit cell $R$ and atom orbital $i$. 

First we analyse the non-interacting part of the Hamiltonian. We are in the dimerised limit meaning that this can be diagonalised using,
\begin{equation}
c^\dagger_{R, \pm} = \frac{1}{\sqrt{2}}\left(c^\dagger_{R, A}\pm c^\dagger_{R, B}\right),
\end{equation}
in this basis,
\begin{equation}
\hat{H}_0 = v\sum_{R} \left(c^\dagger_{R, +} c_{R, +} -  c^\dagger_{R, -} c_{R, -}\right).
\end{equation}

\subsection{Conserved quantities}
This model has an extensive number of conserved quantities which allow us to find the exact solution. Defining the number operator for each dimer as $\hat n_{R} = \hat n_{R, A}+\hat n_{R ,B}$ then, 
\begin{equation}
[\hat H, \hat n_{R}] = 0 \:\:\forall R.
\end{equation}
Intuitively this is because the electrons cannot hop out of each unit cell (as we are in the dimerised limit). Therefore the number of electrons per unit cell is a conserved quantity. 

The Hamiltonian therefore separates into blocks which have different numbers of electrons in each dimer. We consider half-filling where the non-interacting model is an insulator. We study the block in which we have one electron per unit cell. We next show that we can exactly solve the Hamiltonian in this block using a mapping to the transverse field Ising model.

\subsection{Spin chain}
First we note that, since we have one electron per unit cell, we can neglect the intraunit cell interaction (as we never have two adjacent electrons \emph{within} a dimer). We wish to rewrite the Hamiltonian in Eq.~\eqref{eq:totalHamiltonian} as a spin Hamiltonian in terms of the raising and lowering operators,
\begin{equation}
\sigma^+_{R} = c^\dagger_{R, +} c_{R, -},\quad \sigma^-_{R} = c^\dagger_{R, -} c_{R, +},
\end{equation}
where $\sigma^{+/-}_{R}$ are the (Pauli) raising and lowering operators for the $R^{\mathrm{th}}$ spin in the chain. Using the definition $\sigma_R^x = \sigma^+_{R} + \sigma^-_{R}$ we find that,
\begin{align}
\sigma_R^x &= \hat n_{R, A} - \hat n_{R, B} \\
&= 2\hat n_{R, A} - 1 \\
&=  1 - 2\hat n_{R, B},
\end{align}
since $\hat n_{R} = \hat n_{R, A} + \hat n_{R, B}  = 1$ in this regime. Finally we note that,
\begin{align}
\sigma^z_{R} &= [\sigma^+_R, \sigma^-_R]\\
&=  c^\dagger_{R, +} c^{\vphantom{\dagger}}_{R, +} -  c^\dagger_{R, -} c^{\vphantom{\dagger}}_{R, -}.
\end{align}
This set of operators has the desired spin algebra so we can rewrite our original Hamiltonian as a spin Hamiltonian,
\begin{align}
\hat H &= v \sum_R \sigma^z_R + \frac{V_{\mathrm{int}}}{4}\sum_R \left(1 - \sigma^x_R\right)\left(\sigma_{R+1}^x+1\right)\\
&= v \sum_R \sigma^z_R - \frac{V_{\mathrm{int}}}{4}\sum_R \sigma_{R}^x \sigma_{R+1}^x \\&\qquad+ \frac{V_{\mathrm{int}}}{4}\sum_{R} (\sigma_{R+1}^x - \sigma_{R}^x) + \frac{V_{\mathrm{int}} L_c}{4}.\nonumber
\end{align}
Since we are in periodic boundary conditions we can simplify this to,
\begin{align}
\hat H &= v \sum_R \sigma^z_R - \frac{V_{\mathrm{int}}}{4}\sum_R \sigma^x_R \sigma_{R+1}^x +\frac{V_{\mathrm{int}} L_c}{4}
\end{align}
This is the transverse field Ising model Hamiltonian. Similarly in open-boundary conditions we obtain a transverse field Ising model but with boundary fields \emph{i.e.} $\pm\frac{V_{\mathrm{int}}}{4}\sigma_{0}^{x}$. The sign of these boundary fields depends on whether the dimers that are adjacent to the chain of single occupied dimers are doubly or singly occupied. See Tab.~\ref{tab:boundary_field_orientations} for prescriptions of the boundary fields and the constant offset energy. We show how they depend on the occupation of the dimers at either end of the chain of singly occupied dimers (of length $L_c$).

\begin{table}[]
    \centering
    \begin{tabular}{|c|c|l|l|} \hline 
         $R = -1$&  $R = L_c$ & Boundary Fields&Offset Energy\\\hline\hline
 0& 0& $\frac{V_{\mathrm{int}}}{4}(-\sigma_{0}^x + \sigma_{L_c-1}^x)$&$\frac{V_{\mathrm{int}}(L_c-1)}{4}$\\\hline  
         1& 
      0& $\frac{V_{\mathrm{int}}}{4}(-\sigma_0^x - \sigma_{L_c-1}^x)$&$\frac{V_{\mathrm{int}}(L_c-1)}{4} + \frac{V_{\mathrm{int}}}{2}$\\ \hline
 0& 1& $\frac{V_{\mathrm{int}}}{4}(\sigma_0^x + \sigma_{L_c-1}^x)$&$\frac{V_{\mathrm{int}}(L_c-1)}{4} + \frac{V_{\mathrm{int}}}{2}$\\\hline
 1& 1& $\frac{V_{\mathrm{int}}}{4}(\sigma_0^x - \sigma_{L_c-1}^x)$&$\frac{V_{\mathrm{int}}(L_c-1)}{4} + V_{\mathrm{int}}$\\\hline\end{tabular}
    \caption{Effect of adding half dimer on the Spin Hamiltonian. The left and right occupation number gives whether the half dimers which terminate the chain on either end are occupied or not. The half-dimer occupations change the boundary fields and the offset energy. }
    \label{tab:boundary_field_orientations}
\end{table}
\newpage
\section{Energy cost of doubly occupied dimers in the dimerised limit}\label{app:doubleOccDimersProof}
In this section we show that adding doubly occupied dimers always costs of order $V_{\mathrm{int}}$. In the dimerised limit the MB spectrum consists of flat bands. Each flat band can be labeled by the number of anti-bonding states occupied $n_{a}$. Since we are at half filling with $L_c$ unit cells we therefore have $L_c - n_{a}$ bonding orbitals occupied. We restrict ourselves to the subspace of one of these MB bands and see how adding interactions changes the energy of the states with different numbers of doubly occupied dimers. 

Consider the sector with $n_d$ doubly occupied dimers with $0\le n_d\le n_{a}$. A doubly occupied dimer and empty dimer pair have lowest energy when they are adjacent. Therefore the lowest energy state in the $n_d$ doubly occupied subspace has all $n_d$ doubly occupied dimer and empty dimer pairs in adjacent unit cells. To lower their energy the unit cells in this region will alternate between doubly occupied and empty. The remaining electrons will lie in one electron per unit cell in the remaining $L_c-2n_d$ unit cells. 

We now calculate the energy of this state. Firstly the region of doubly occupied and empty dimers contributes $V_{\mathrm{int}} n_d$ energy. The remaining dimers can be mapped to a transverse field Ising model since there is one electron per dimer. However, we have the chain in open-boundary conditions and so we have some boundary fields (see Tab.~\ref{tab:boundary_field_orientations}). The transverse field Ising model Hamiltonian for this chain will therefore be,
\begin{align}
\hat H &= v\sum_{R = 0}^{L_c - 2n_d-1} \sigma^z_R -\frac{V_{\mathrm{int}}}{4} \sum_{R = 0}^{L_c - 2n_d-2}\sigma^x_R \sigma^x_{R+1} \\&\quad+ \frac{V_{\mathrm{int}}}{4} \sigma^x_0+ \frac{V_{\mathrm{int}}}{4} \sigma^x_{L_c - 2n_d-1} + \frac{V_{\mathrm{int}}}{4}(L_c - 2n_d +1).\nonumber\end{align}
(Recall that we defined $v \equiv 1+ \delta$). We now note that we are restricting our attention to one $n_b$ sector. (Recall $n_b$ labels the number of bonding states in the band). Therefore we can restrict our Hamiltonian to operators which conserve the number of bonding states. The $\sigma^z_R$ is diagonal in the basis of bonding/anti-bonding states but the $\sigma^x_R$ can flip bonding/anti-bonding states. Therefore we rewrite terms in terms of the raising and lowering operators where $\sigma_{R}^x = \sigma_R^++ \sigma^-_R$. This means that the boundary fields become $\sigma^x_0 = \sigma_0^++ \sigma^-_0 $. This term therefore maps us out of the subspace with $n$ bonding states. We can therefore drop this term. The term Ising coupling is $\sigma^x_R \sigma^x_{R+1} = (\sigma_R^++ \sigma^-_R)(\sigma_{R+1}^++ \sigma^-_{R+1})$. We can only allow terms which hop the bonding state rather than create or destroy a bonding and anti-bonding state pair. Therefore this term becomes $\sigma^x_R \sigma^x_{R+1} \approx \sigma_R^+ \sigma^-_{R+1}+\sigma_R^- \sigma^+_{R+1}$. Our model becomes,
\begin{align}
\hat H_{n} &= v\sum_{R = 0}^{L_c - 2n_d-1} \sigma^z_R -\frac{V_{\mathrm{int}}}{4} \sum_{R = 0}^{L_c - 2n_d-2}(\sigma_R^+ \sigma^-_{R+1}+\sigma_R^- \sigma^+_{R+1}) \nonumber\\&\quad+ \frac{V_{\mathrm{int}}}{4}(L_c - 2n_d +1).\end{align}
This can be mapped to a simple non-interacting fermion chain by identifying $\sigma^+_R \equiv c^\dagger_R$, $\sigma^-_R \equiv c_R$ and $c^\dagger_{R}c_{R} - c_Rc^\dagger_R\equiv \sigma_R^z$. Therefore the Hamiltonian can be rewritten as,
\begin{align}
\hat H_{n} &= 2v\sum_{R = 0}^{L_c - 2n_d-1} c^\dagger_R c_R -\frac{V_{\mathrm{int}}}{4} \sum_{R = 0}^{L_c - 2n_d-2}(c^\dagger_R c_{R+1} + c^\dagger_{R+1} c_{R}) \nonumber\\&\quad+ \frac{V_{\mathrm{int}}}{4}(L_c - 2n_d +1) - v(L_c - 2 n_d).\label{eq:tightbindingTFIM}\end{align}
For now we drop the constant energy shift to give the Hamiltonian,
\begin{equation}
\hat H_{n}' = 2v\sum_{R = 0}^{L_c - 2n_d-1} c^\dagger_R c_R -\frac{V_{\mathrm{int}}}{4} \sum_{R = 0}^{L_c - 2n_d-2}(c^\dagger_R c_{R+1} + c^\dagger_{R+1} c_{R}).
\end{equation}
The spectrum of this is,
\begin{equation}
E_l' = 2v - \frac{V_{\mathrm{int}}}{2}\cos\left(\frac{l\pi}{L_c - 2n_d+1}\right),
\end{equation}
for $l\in \mathbb{Z}$. This represents the energy cost of adding one anti-bonding dimer on top of a ground state of bonding orbitals. We see that the energy cost of adding a anti-bonding orbital is $\ge 2v - \frac{V_{\mathrm{int}}}{2}$. 

In our case, we have $n_d$ anti-bonding dimers in doubly occupied dimers. So there are $n_{a} - n_d$ anti-bonding dimers in the region with one electron per dimer. The energy contribution of these is lower bounded by $(n_{a} - n_d)(2v - \frac{V_{\mathrm{int}}}{2})$ in addition to the constant energy offset in Eq.~\eqref{eq:tightbindingTFIM}. Therefore we can write down a lower bound for the total energy of the lowest energy state in the MB band (labeled by $n_a$) in the $n_d$ sector. This is,
\begin{equation}
E_{n_b, n_d} = 2v\left(n_{a} - \frac{L_c}{2}\right) + \frac{V_{\mathrm{int}}}{2}\left(2n_d - n_a+\frac{L_c}{2}+\frac{1}{2}\right)
\end{equation}
We see that adding a single extra doubly occupied dimer adds approximately $V_{\mathrm{int}}$ in energy as expected. This justifies the $V_{\mathrm{int}}$ in the denominator of the matrix element in Eq.~\eqref{eq:dimerizedMatrixElement}. By a similar calculation we can show that the highest energy states in the $n_d$ and $n_d+1$ sectors also differ by approximately $V_{\mathrm{int}}$.

We therefore conclude that the highest energy state within the band with $n_a$ anti-bonding states has the maximum number of doubly occupied dimers (\textit{i.e.} $n_a$). Similarly the lowest energy state within the band has no doubly occupied dimers. Lastly the sectors with different $n_d$ doubly occupied dimers and $n_d+1$ differ in energy by of order $V_{\mathrm{int}}$.

\newpage
\section{Interaction in the band basis for the SSH model}
\label{app:selectionrule}
 \begin{figure*}
     \centering
    \includegraphics[width=0.49\linewidth]{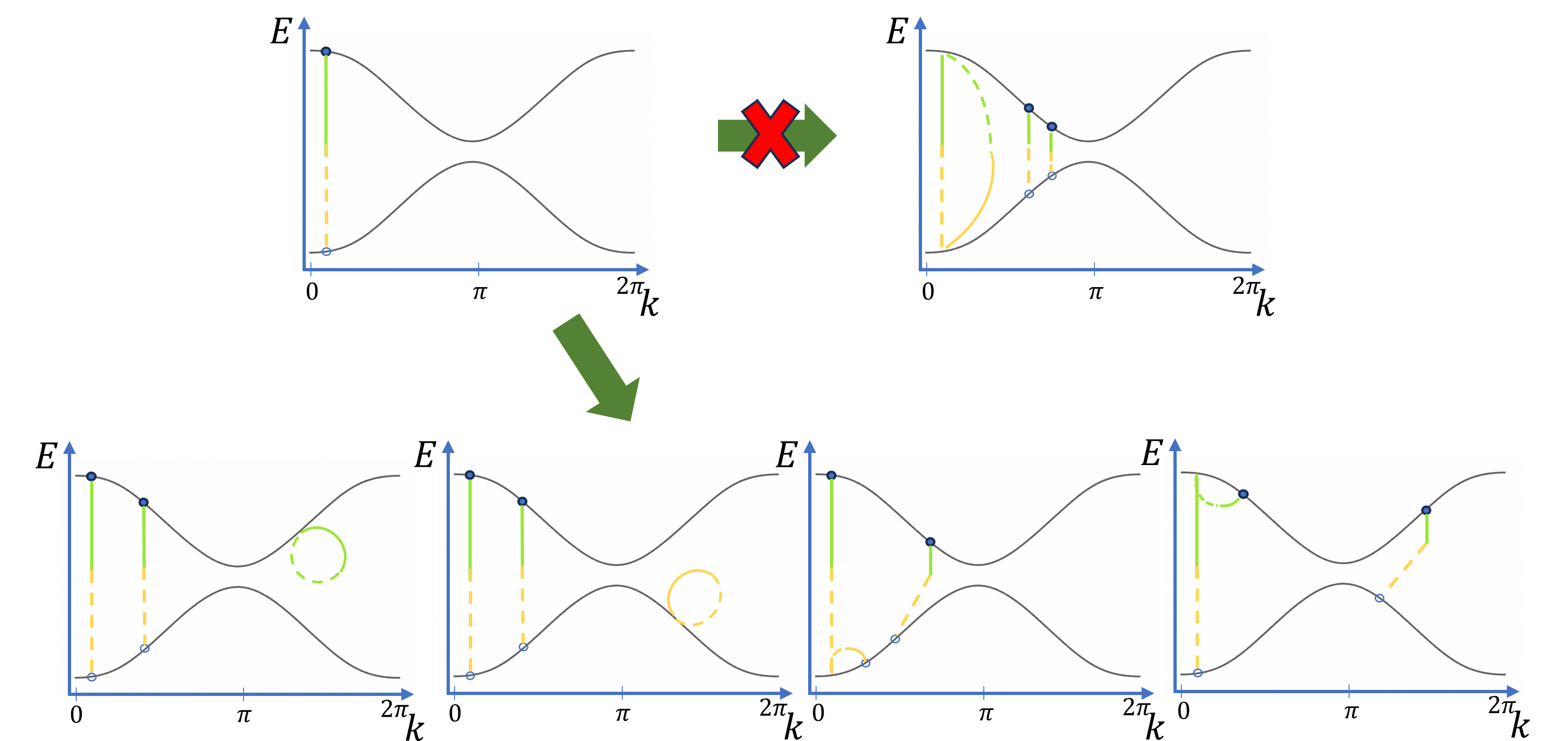}
    \includegraphics[width=0.49\linewidth]{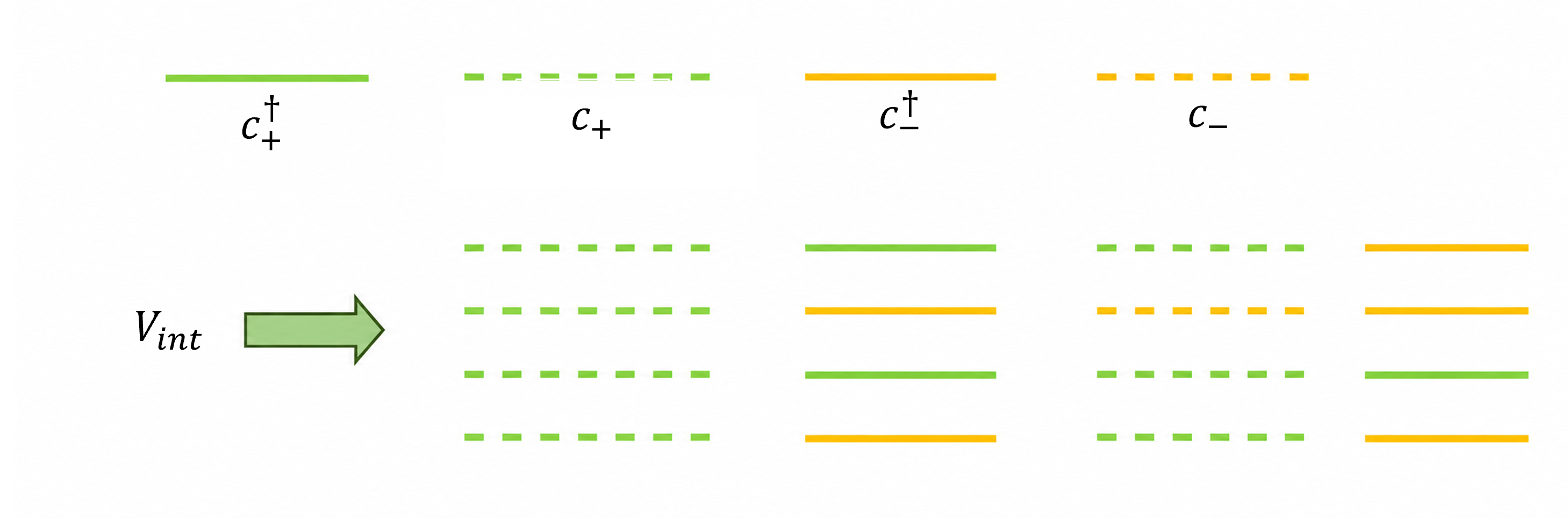}
     \caption{Schematic illustration of miniband structure and interaction-induced scattering processes. \textbf{Top left:} A representative state in the first excited miniband near the top, obtained by promoting a particle from the filled lower band to the upper band (dashed/solid line indicates the excitation process). \textbf{Top right:} A representative state near the bottom of the second miniband, involving two such excitations, which cannot be reached directly via $V_{\rm int}$.
\textbf{Bottom:} Scattering processes generated by $V_{\rm int}$. Allowed processes involve two annihilation and two creation operators and can transfer particles between bands while conserving total momentum. Crucially, these processes cannot redistribute both the particle and hole parts of an already existing excitation to a different momentum while simultaneously creating a new one.
\textbf{Right:} Legend for the notations above: dashed (solid) lines denote creation (annihilation) processes. Colors distinguish operators acting in the upper ($c_{+}$) and lower ($c_{-}$) bands. Example vertices illustrate allowed two-body interaction processes consistent with allowed constraints.}
     \label{fig:bandstructure}
 \end{figure*}

In this appendix, we provide a detailed decomposition of the nearest-neighbour density-density interaction in the band basis of the Su-Schrieffer-Heeger (SSH) model, and classify all resulting processes according to their action on the half-filled ground state and on excited states.

\subsection{Diagonalized SSH Hamiltonian}

The SSH Hamiltonian in momentum space reads
\begin{equation}
H = \sum_k \begin{pmatrix}
c^\dagger_{k,A} & c^\dagger_{k,B}
\end{pmatrix}
\begin{pmatrix}
0 & h(k) \\
h^*(k) & 0
\end{pmatrix}
\begin{pmatrix}
c_{k,A} \\
c_{k,B}
\end{pmatrix},
\end{equation}
where $h(k) = (1+\delta) + (1-\delta) e^{-ik} = |h(k)| e^{-i\phi_k}$ and $c^\dagger_{k, i} = \frac{1}{\sqrt{L}}\sum_{R} e^{\mathrm{i}kR} c^\dagger_{R, i}$

The Hamiltonian is diagonalized via
\begin{equation}
\begin{pmatrix}
c_{k,A} \\
c_{k,B}
\end{pmatrix}
=\frac{1}{\sqrt{2}}
\begin{pmatrix}
1 & 1 \\

-e^{i\phi_k} & e^{i\phi_k}
  \end{pmatrix}
  \begin{pmatrix}
  c_{k,-} \\
  c_{k,+}
  \end{pmatrix},
  \label{eq:U}
  \end{equation}
  leading to
  \begin{equation}
  H = \sum_k \left( -|h(k)| c_{k,-}^\dagger c_{k,-} + |h(k)| c_{k,+}^\dagger c_{k,+} \right).
  \end{equation}

\subsection{Ground state and excitations}

At half filling, the ground state is
\begin{equation}
|\mathrm{GS}\rangle = \prod_k c_{k,-}^\dagger |0\rangle,
\end{equation}
where all lower-band states are filled and upper-band states are empty.

A single particle--hole excitation is created by
\begin{equation}
|k\rangle = c_{k,+}^\dagger c_{k,-} |\mathrm{GS}\rangle.
\end{equation}

\subsection{Interaction Hamiltonian}

We consider a nearest-neighbour interaction
\begin{equation}
\hat{H}_{\mathrm{int}} = V \sum_i n_i n_{i+1},
\end{equation}
which in momentum space takes the schematic form
\begin{equation}
\hat{H}_{\mathrm{int}} = \sum V_{k_1 k_2 k_3 k_4} c^\dagger_{k_1,A} c_{k_2,A} c^\dagger_{k_3,B} c_{k_4,B},
\end{equation}
with momentum conservation requiring $k_1-k_2+k_3-k_4=0$.

Transforming to the band basis, we express the sublattice operators in terms of the eigenmodes of the Bloch Hamiltonian. Writing the diagonalization in component form,
\begin{equation}
c_{k,\alpha} = \sum_{\beta = +,-} U_{\alpha \beta}(k)\, c_{k,\beta}, 
\qquad \alpha \in \{A,B\},
\end{equation}
where $U_k$ is the unitary matrix that diagonalizes the SSH Hamiltonian, and $\mu_k = c_{k,-}, c_{k,+}$ denote annihilation operators in the lower and upper bands respectively.

From Eq.~\eqref{eq:U}, we have explicitly
\begin{equation}
U_k = \frac{1}{\sqrt{2}}
\begin{pmatrix}
1 & 1 \\
- e^{i\phi_k} & e^{i\phi_k}
\end{pmatrix},
\end{equation}
which gives
\begin{equation}
c_{k,A} = \frac{1}{\sqrt{2}}(c_{k,-} + c_{k,+}), \qquad
c_{k,B} = \frac{e^{i\phi_k}}{\sqrt{2}}(-c_{k,-} + c_{k,+}).
\end{equation}
\begin{equation}
\hat{H}_{\mathrm{int}} \sim \sum (\mu^\dagger \nu)(\rho^\dagger \sigma), \quad \mu,\nu,\rho,\sigma \in \{c_{k,+},c_{k,-}\},
\end{equation}
which generates $2^4 = 16$ operator structures.

\subsection{Classification of interaction terms}

The sixteen quartic operator structures can be classified according to their effect on the number of particle--hole excitations. Writing each term schematically as $(\mu^\dagger \nu)(\rho^\dagger \sigma)$ with $\mu,\nu,\rho,\sigma \in \{c_{-},c_{+}\}$, we drop the index $k$ at the moment for simplicity and group them as follows:

\paragraph{Density terms (no excitation)}

\begin{align}
& c_{-}^\dagger c_{-}\, c_{-}^\dagger c_{-}, \\
& c_{-}^\dagger c_{-}\, c_{+}^\dagger c_{+}, \\
& c_{+}^\dagger c_{+}\, c_{-}^\dagger c_{-}, \\
& c_{+}^\dagger c_{+}\, c_{+}^\dagger c_{+}.
\end{align}

These preserve band occupations and act diagonally. On the ground state they reduce to c-number contributions.

\paragraph{Exchange (scattering) terms (no net excitation)}

\begin{align}
& c_{-}^\dagger c_{+}\, c_{+}^\dagger c_{-}, \\
& c_{+}^\dagger c_{-}\, c_{-}^\dagger c_{+},
\end{align}

These conserve the number of excitations. The first of the two acts diagonally on the ground state while the second vanishes on the ground state. Both however, act non-trivially within excited sectors, describing scattering processes.

\paragraph{Single particle--hole creation}

\begin{align}
& c_{+}^\dagger c_{-}\, c_{-}^\dagger c_{-}, \\
& c_{-}^\dagger c_{-}\, c_{+}^\dagger c_{-}, \\
& c_{+}^\dagger c_{-}\, c_{+}^\dagger c_{+}, \\
& c_{+}^\dagger c_{+}\, c_{+}^\dagger c_{-}.
\end{align}

These contain one free $c_{+}^\dagger$ and one free $c_{-}$. The third case vanishes on action on ground state due to $c_{+}^\dagger c_{+}  |\mathrm{GS}\rangle = 0$.

\paragraph{Single particle--hole annihilation (de-excitation)}

\begin{align}
& c_{-}^\dagger c_{+} \, c_{-}^\dagger c_{-}, \\
& c_{-}^\dagger c_{-}\, c_{-}^\dagger c_{+}, \\
& c_{-}^\dagger c_{+}\, c_{+}^\dagger c_{+}, \\
& c_{+}^\dagger c_{+}\, c_{-}^\dagger c_{+}.
\end{align}

These remove an excitation from the $+$ band and create a particle in the $-$ band. Thus all of them vanish on application on the Ground state. However, all of them give non zero contribution on the first excited state, apart from the last case which only contributes from the second excitation sector. 

\paragraph{Two particle-hole creation}

\begin{align}
& c_{+}^\dagger c_{-}\, c_{+}^\dagger c_{-}.
\end{align}

\paragraph{Two particle-hole annihilation}

\begin{align}
& c_{-}^\dagger c_{+}\, c_{-}^\dagger c_{+}.
\end{align}

These annihilate both the ground state and any single-excitation state.

\subsection{Examples of action on a state in the $n$-excitation sector}

Let us explicitly enumerate what each term discussed above does to this state
\begin{equation}
|k_1,k_2,\hdots, k_n\rangle = \prod_{i=1}^{n} c_{k_i,+}^\dagger c_{k_i,-} |\mathrm{GS}\rangle.
\end{equation}

Using
\begin{equation}
c_{k,-}^\dagger |\mathrm{GS}\rangle = 0, \quad c_{k,+} |\mathrm{GS}\rangle = 0,
\end{equation}
we classify:

\paragraph{ No-excitation  terms}

To make the action precise, we restore momentum labels.

This sector corresponds to choosing identical band indices within each bilinear and depending on the momentum labels can just compute the density-density correlation or scatter particle holes to different momentum. We show an example for the latter.
\begin{align}
& c_{q_1,-}^\dagger c_{q_2,-}\, c_{q_3,+}^\dagger c_{q_4,+},
\end{align}
with momentum conservation providing the constraint $q_1 - q_2 + q_3 - q_4 = 0$.

Let us study the action of this term on the typical state of a $n$-excitation sector.
We start from a half- filled ground state
\begin{equation}
|\mathrm{GS}\rangle = \prod_k c_{k,-}^\dagger |0\rangle;
\end{equation}

We use fermionic anticommutation relations
\begin{equation}
c_{k,\alpha} c^\dagger_{q,\beta} = \delta_{kq}\delta_{\alpha \beta } - c^\dagger_{q,\beta} c_{k,\alpha},
\end{equation}
where $\alpha,\beta \in \{+,-\}$, to arrange the terms in the correct normal ordering, bringing $c_{k,+}$ and $c_{k,-}^{\dagger}$ to the extreme right. This provides us with the Wick contractions.

The action of the upper-band bilinear gives
\begin{align}
c_{q_3,+}^{\dagger}c_{q_4,+}
\prod_{i=1}^{n}c_{k_i,+}^{\dagger}
&=
\sum_{j=1}^{n}
(-1)^{j-1}
\delta_{q_4,k_j}
c_{q_3,+}^{\dagger}
\prod_{i\neq j}c_{k_i,+}^{\dagger} + c_{q_3,+}^{\dagger}\prod_{i=1}^n c_{k_i,+}^{\dagger}c_{q_4,+},
\end{align}
where the last term vanishes upon application on the ground state since there are no additional occupied states in the upper band. This corresponds to removing an excitation at momentum $q_4$ and creating one at momentum $q_3$.

\begin{align}
c_{q_1,-}^{\dagger}c_{q_2,-}
\prod_{i=1}^{n}c_{k_i,-}
|\mathrm{GS}\rangle
&=
\sum_{j=1}^{n}
(-1)^{j-1}
\delta_{q_1,k_j}
\prod_{i\neq j}c_{k_i,-}
c_{q_2,-}
|\mathrm{GS}\rangle
\nonumber\\
&\quad+
(-1)^n
\prod_{i=1}^{n}c_{k_i,-}
c_{q_1,-}^{\dagger}c_{q_2,-}
|\mathrm{GS}\rangle .
\end{align}

The first term corresponds to refilling an existing hole at momentum $q_1$ while creating a new hole at momentum $q_2$, whereas the second term contributes only when $q_1=q_2$, giving a diagonal density contribution.

\paragraph{Single particle-hole creation}

We now analyze interaction terms that create exactly one additional particle-hole excitation on top of a state containing \(n\) particle-hole pairs. A representative operator is

\begin{equation}
c^\dagger_{q_1,+} c_{q_2,-}\, c^\dagger_{q_3,-} c_{q_4,-},
\end{equation}
subject to momentum conservation
\(
q_1-q_2+q_3-q_4=0.
\)

We first evaluate the action of the lower-band operators by commuting 
\(c^\dagger_{q_3,-}\) through the existing hole operators:
\begin{align}
c^\dagger_{q_3,-}c_{q_4,-}
\prod_{i=1}^{n}c_{k_i,-}
|\mathrm{GS}\rangle
&=
\sum_{j=1}^{n}
(-1)^{j-1}
\delta_{q_3,k_j}
\prod_{i\neq j}c_{k_i,-}
c_{q_4,-}
|\mathrm{GS}\rangle
\nonumber\\
&\quad+
(-1)^n
c_{q_4,-}
\prod_{i=1}^{n}c_{k_i,-}
c^\dagger_{q_3,-}
|\mathrm{GS}\rangle .
\end{align}

The first term corresponds to refilling an existing hole at momentum \(q_3\) while creating a new hole at momentum \(q_4\). The second term is normal ordered and therefore annihilates the ground state.

The operator \(c_{q_2,-}\) removes an additional fermion from the filled lower band, creating another hole excitation, while \(c^\dagger_{q_1,+}\) creates a particle in the upper band. Consequently, the total number of particle-hole excitations changes as

\begin{equation}
n\rightarrow n+1.
\end{equation}

Importantly, such processes cannot simultaneously reshuffle both constituents of an existing particle-hole pair while creating an additional one. This follows directly from the quartic structure of the interaction: two operators create fermions and two annihilate them, with three operators acting within one band and only one acting in the other. After creating the additional particle-hole pair, at most one operator can act non-trivially on the original excitation manifold, see Fig.~\ref{fig:bandstructure} for a visual representation.

For the \(n=1\) sector, this implies that either

\begin{itemize}
    \item the original particle is scattered while the hole remains unchanged with a new pair creation,
    \item the original hole is scattered while the particle remains unchanged with a new pair creation, or
    \item the original particle-hole pair remains unchanged while a new pair is created.
\end{itemize}

Thus, at least one constituent of the original excitation necessarily remains intact, implying that the final state always retains an excitation at the original single-particle energy scale. The same argument extends straightforwardly to arbitrary excitation sectors \(n=r\).

This immediately constrains the energetics of states coupled by \(V_{\rm int}\). For a transition from the \(n\)-excitation sector to the \((n+1)\)-excitation sector, the energy difference is

\begin{equation}
\Delta E=
\epsilon(q_1)+\epsilon(q_2)+\epsilon(q_3)+\epsilon(q_4)-2\epsilon(k).
\end{equation}

From the discussion above, one of the momenta must satisfy \(q_4=k\), and momentum conservation further imposes

\begin{equation}
q_1-q_2+q_3=k.
\end{equation}

This reduces the energy mismatch to

\begin{equation}
\Delta E=
\epsilon(q_1)+\epsilon(q_2)+\epsilon(q_3)
-\epsilon(q_1-q_2+q_3).
\end{equation}

Using Eq.~\eqref{eq:singleparticledispersion}, the single-particle spectrum satisfies

\begin{equation}
2\delta \leq \epsilon(q)\leq 2,
\qquad 0<\delta<1.
\end{equation}

This immediately gives the bound

\begin{equation}
\Delta E \geq 6\delta-2,
\end{equation}

which guarantees \(\Delta E\geq 0\) for \(\delta>1/3\). Numerically, however, we find the stronger result

\begin{equation}
\Delta E>0,\qquad \forall\, \delta>0.
\end{equation}

A rigorous proof of this stronger bound is left for future work. Nevertheless, this establishes that for any finite \(\delta\), the interaction cannot resonantly couple physical states and can only contribute through virtual intermediate states.

\paragraph{Two particle-hole creation:} Finally, processes that create two particle-hole excitations act independently and do not involve scattering of existing excitations, which automatically provides the same conclusion about absence of resonant coupling.

\newpage
\section{Higher-order hybridization paths between minibands}

\label{app:pert}
\begin{figure}
    \centering
    \includegraphics[width=0.9\linewidth]{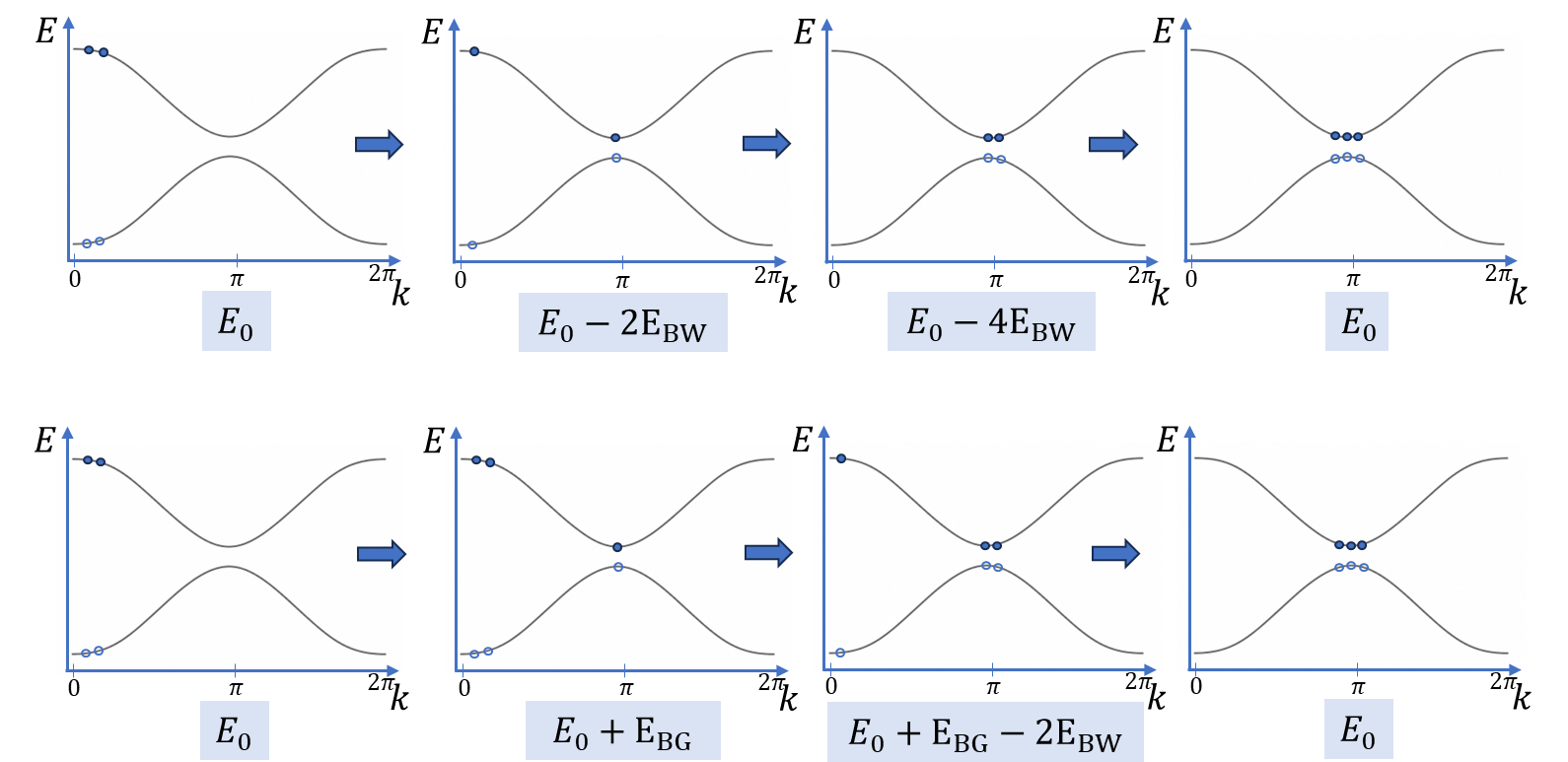}
    \caption{Perturbative pathways}
    \label{fig:pathways}
\end{figure}

In this appendix, we explicitly illustrate why spectral overlap between many-body minibands does not immediately imply strong hybridization. As discussed in Appendix~\ref{app:selectionrule}, the interaction term can create at most one additional particle-hole excitation while simultaneously scattering only one existing excitation. Consequently, certain states that become degenerate across minibands cannot hybridize at low order.

To make this explicit, we consider hybridization between a high-energy state in the $m=2$ miniband and a low-energy state in the $m=3$ miniband, shown schematically in Fig.~\ref{fig:pathways}. We focus on the regime where these two states are energetically degenerate,
\begin{equation}
E_i=E_f=E_0,
\end{equation}
which corresponds to the onset of overlap between these minibands.

The initial state corresponds to two excitations near the \emph{top} of the upper single-particle band ($k\approx0$),
\begin{equation}
|i\rangle \sim 
c^\dagger_{0,+}c_{0,-}\,
c^\dagger_{0,+}c_{0,-}|{\rm GS}\rangle,
\end{equation}
while the final state contains three excitations near the \emph{bottom} of the upper band ($k\approx\pi$),
\begin{equation}
|f\rangle \sim 
c^\dagger_{\pi,+}c_{\pi,-}\,
c^\dagger_{\pi,+}c_{\pi,-}\,
c^\dagger_{\pi,+}c_{\pi,-}|{\rm GS}\rangle.
\end{equation}

Since the interaction cannot simultaneously create one excitation and scatter both existing excitations, the transition necessarily proceeds through multiple virtual steps. Figure~\ref{fig:pathways} illustrates two extremal possibilities.

\paragraph{Path I: scatter first, create later}

In the first process, the two existing excitations are sequentially scattered from the top of the band toward the band minimum before the final excitation is created:
\begin{align}
|i\rangle 
\rightarrow 
|n_1\rangle 
\rightarrow 
|n_2\rangle 
\rightarrow 
|n_3\rangle 
\rightarrow 
|f\rangle.
\end{align}

The corresponding intermediate energies scale as
\begin{align}
E_{n_1} &\approx E_i-2E_{\rm BW},\\
E_{n_2} &\approx E_i-4E_{\rm BW},\\
E_{n_3} &\approx E_i,
\end{align}
where
\begin{equation}
E_{\rm BW}=E_{\max}-E_{\min}
\end{equation}
is the single-particle bandwidth.

This corresponds to the top row of Fig.~\ref{fig:pathways}.

\paragraph{Path II: create first, scatter later}

Alternatively, the interaction may first create an additional excitation and subsequently scatter all excitations toward the band minimum:
\begin{align}
|i\rangle 
\rightarrow 
|m_1\rangle 
\rightarrow 
|m_2\rangle 
\rightarrow 
|m_3\rangle 
\rightarrow 
|f\rangle.
\end{align}

The intermediate energies now scale as
\begin{align}
E_{m_1} &\approx E_i+E_{\rm BG},\\
E_{m_2} &\approx E_i+E_{\rm BG}-2E_{\rm BW},\\
E_{m_3} &\approx E_i,
\end{align}
where
\begin{equation}
E_{\rm BG}=2E_{\min}
\end{equation}
denotes the many-body excitation gap associated with creating an additional particle-hole pair.

This corresponds to the bottom row of Fig.~\ref{fig:pathways}.

Although the precise intermediate energies depend on the microscopic momentum configurations, both classes of processes require virtual excursions controlled parametrically by $E_{\rm BW}$ and $E_{\rm BG}$.

\paragraph{Van Vleck effective coupling}

Projecting onto the degenerate manifold using Van Vleck perturbation theory, the effective matrix element between the initial and final states takes the form

\begin{equation}
H_{\rm eff}^{(3)}(f,i)=
\sum_{n_1,n_2}
\frac{
V_{fn_2}
V_{n_2n_1}
V_{n_1i}
}{
(E_i-E_{n_1})
(E_i-E_{n_2})
},
\end{equation}
where the sum runs over all allowed intermediate states.

For the two representative paths shown in Fig.~\ref{fig:pathways}, this gives the parametric scaling

\begin{equation}
t_{\rm eff}^{(3)}
\sim
\frac{V_{\rm int}^3}{(2E_{\rm BW})(4E_{\rm BW})}
+
\frac{V_{\rm int}^3}
{E_{\rm BG}(E_{\rm BG}-2E_{\rm BW})}.
\end{equation}

These are only two representative paths; in practice many such virtual processes contribute.

\paragraph{General $n$-step processes}

More generally, for states separated by $n$ constrained operations, the effective coupling takes the form

\begin{equation}
H_{\rm eff}^{(n)}(f,i)=
\sum_{\{n_\alpha\}}
\frac{
\prod_{\alpha=0}^{n-1}
V_{n_{\alpha+1}n_\alpha}
}{
\prod_{\alpha=1}^{n-1}
(E_i-E_{n_\alpha})
},
\end{equation}
with
\[
n_0=i,\qquad n_n=f.
\]

While the number of allowed paths grows rapidly with perturbative order, each path is suppressed by increasingly large energy denominators. Consequently, miniband overlap alone does not immediately produce strong hybridization. Strong level repulsion emerges only when the proliferation of such paths compensates for this perturbative suppression.
\newpage
\section{Open boundary condition results}

\begin{figure}
    \centering
    \includegraphics[width=0.6\linewidth]{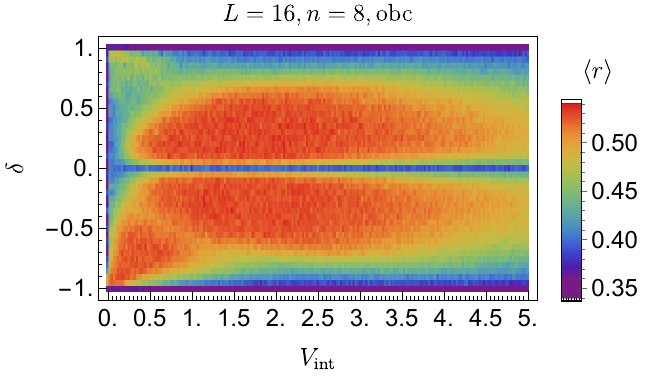}
    \caption{$\langle r \rangle$ statistics for the SSH model under open boundary conditions}
    \label{fig:obc}
\end{figure}

In this appendix, we present the behavior of $\langle r\rangle$ across different parameter regimes for the interacting SSH model with open boundary conditions. In the absence of translation symmetry, the Hilbert space is significantly larger, restricting our numerical simulations to system sizes up to $L=16$ at half filling ($n=8$). The corresponding results are shown in Fig.~\ref{fig:obc}.

Several features are immediately apparent. First, there is a clear qualitative distinction between the topological phase ($\delta>0$) and the trivial phase ($\delta<0$). The trivial phase closely reproduces the behavior observed for periodic boundary conditions. In contrast, in the topological phase, the anomaly is replaced by a more pronounced suppression of $\langle r \rangle$ near the integrable limits. Instead of a well-defined line of low $\langle r \rangle$, one observes an extended region of suppressed $\langle r \rangle$ that persists until sufficiently far from the integrable lines.

Our preliminary analysis suggests that this behavior is related to the presence of topologically protected edge modes in the non-interacting spectrum. These modes hinder efficient hybridization and delay the onset of ergodic behavior, leading to a slower crossover of both level statistics and IPR toward GOE values. A more detailed investigation of this regime is left for future work.
\newpage
\section{Breaking chiral symmetry}\label{app:SymmetryBreakingVariant}
To test the robustness of the mechanism, we explicitly break chiral symmetry by introducing an alternating on-site potential $h$ to the interacting SSH model of Eq.~\eqref{eq:genSSH}, which deforms the SSH single particle band structure while preserving its two-band character. The corresponding single-particle Hamiltonian may be written as $H(k)=\mathbf{d}(k)\cdot\boldsymbol{\sigma}$ with $\mathbf{d}(k)=(t_1+t_2\cos k,\, t_2\sin k,\, h)$. The real-space Hamiltonian is then given by
\begin{eqnarray}
H &=& t_1 \sum_{i\in \text{even}} \left(c_i^\dagger c_{i+1} + \text{H.c.}\right) +
t_2 \sum_{i\in \text{odd}} \left(c_i^\dagger c_{i+1} + \text{H.c.}\right) \nonumber \\
&& + V_{\rm int} \sum_i n_i n_{i+1} + h \sum_i (-1)^i n_i .
\label{eq:chiralbrokenSSH}
\end{eqnarray}

For $h\neq 0$, the chiral symmetry is broken. For these systems, we resolve translation symmetry in the usual manner. In general ($t_1 \neq t_2$), the model does not possess reflection (parity) symmetry, as it is broken by the alternating hopping amplitudes and the staggered potential. In the special case $t_1 = t_2$, parity symmetry is restored, and additional symmetry resolution is possible.

The breaking of chiral symmetry also removes the exact particle-hole symmetry of the model. Instead, at half filling, the system exhibits a particle-hole-like symmetry defined by
\begin{equation}
C\, c_j\, C^\dagger = (-1)^j c_{j+1},
\end{equation}
where $C$ denotes the corresponding symmetry operator~\cite{PhysRevResearch.4.033119}.

Rather than explicitly resolving this symmetry in the numerical implementation, we work away from half filling, where it is automatically broken. The absence of these symmetries increases the effective Hilbert space size. Consequently, we consider system sizes $L=22$ with $n=7$ particles for $t_1 \neq t_2$, and $L=24$ with $n=7$ particles for $t_1 = t_2$, where the enhanced symmetry allows for a finer decomposition into momentum sectors.

For $h\neq0$ the associated integrable line at $\delta=1$ is removed, although in the non-interacting limit ($V_{\rm int}=0$), the Slater-determinant structure remains. For finite $h$, the dimer-based splitting is replaced by a sublattice polarization gap, but as long as a finite single-particle gap persists, the MB spectrum continues to organize into a ladder labeled by the number of particles in the upper band.

If the anomalous suppression of $\langle r\rangle$ originates from the incomplete hybridization mechanism discussed above, the intercept of the anomalous line should shift predictably under $(t_1,t_2,h)$ deformations. Calculating the locus where the $m=1$ and $m=3$ minibands merge yields the analytic condition
\begin{equation}
t_1^2 - \frac{5}{2}t_1 t_2 + t_2^2 + h^2 = 0,
\label{eq:bandcondition}
\end{equation}
which determines the intercept in the noninteracting limit. While a controlled analytic expression for the full slope requires incorporating interaction-induced hybridization non-perturbatively, Eq.~\eqref{eq:bandcondition} correctly captures the intercept.

\begin{figure} \centering \includegraphics[width=0.48\columnwidth]{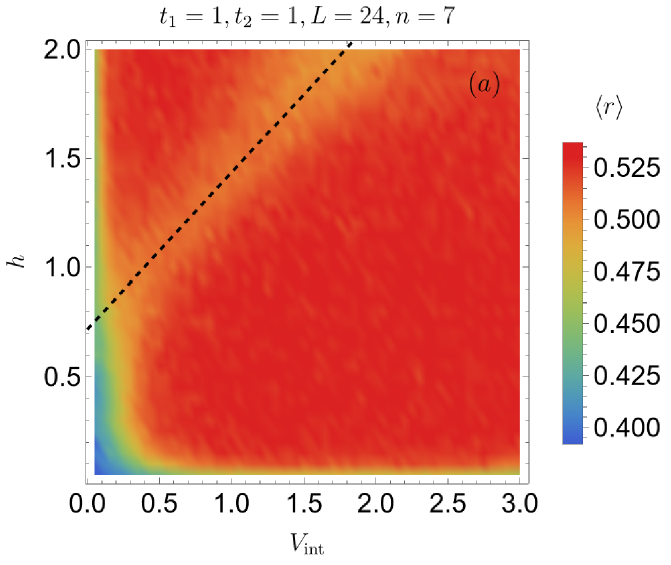} \includegraphics[width=0.48\columnwidth]{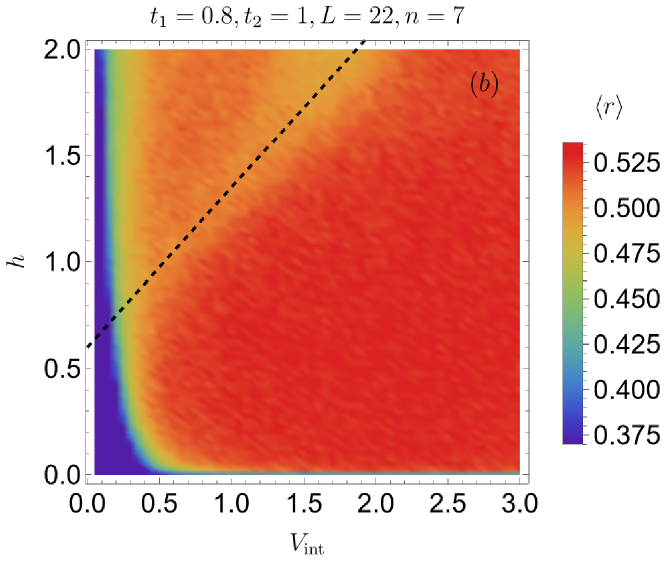}\\ \includegraphics[width=0.48\columnwidth]{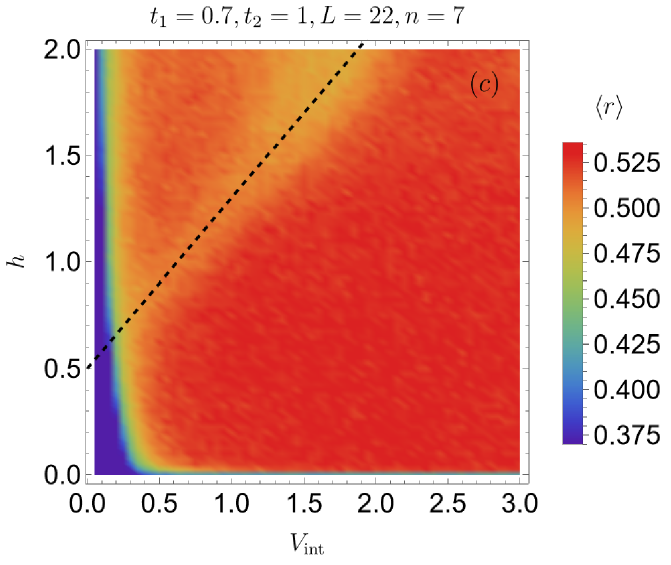} \includegraphics[width=0.48\columnwidth]{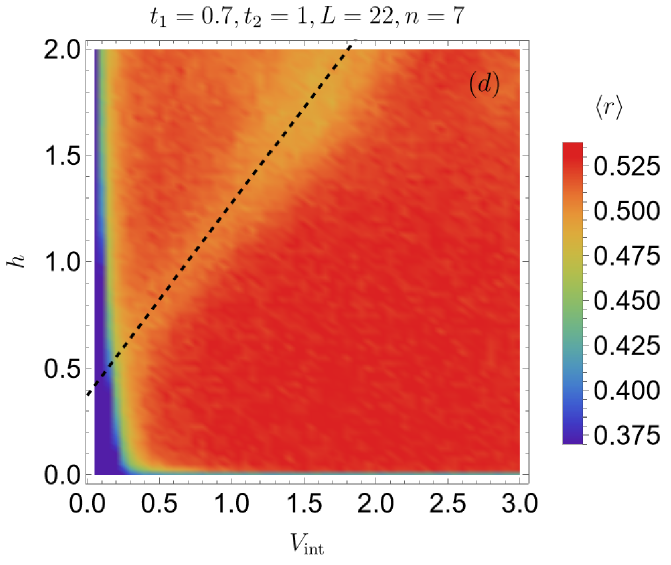} \caption{$\langle r \rangle$ statistics for the model in Eq.~\eqref{eq:genSSH} under varying parameters. Top left: $t_1 = t_2 = 1$, corresponding to uniform hopping with an alternating on-site potential of strength $h$. From Eq.\eqref{eq:bandcondition}, the predicted intercept occurs at $h \sim 1/\sqrt{2} \approx 0.707$, which aligns well with the location of the observed dip in $\langle r \rangle$. Top right: $t_1 = 0.8$; the predicted intercept at $h = 0.6$ again matches the fitted line of slightly lower $\langle r \rangle$. Bottom left: $t_1 = 0.7$ with intercept at $h \approx 0.51$. Bottom right: $t_1 = 0.6$ with intercept at $h \approx 0.37$. In all cases, the analytic prediction from Eq.\eqref{eq:bandcondition} accurately tracks the region of reduced $\langle r \rangle$, suggesting a robust connection between the single-particle spectral structure and emergent MB level statistics. The full line is drawn via a best fit through the middle of low $\langle r \rangle$ region.} \label{fig:onsitetuning} \end{figure}

Figure~\ref{fig:onsitetuning} confirms that the intercept of the anomalous line of reduced $\langle r\rangle$ tracks Eq.~\eqref{eq:bandcondition} across all parameter sets shown. As $h$ is varied at fixed $(t_1,t_2)$, the dip shifts along a linear curve, with the slope depending on $(t_1,t_2)$, demonstrating that the onset of the anomalous regime is controlled by the underlying single-particle spectral geometry rather than by chiral symmetry itself. We see the most pronounced effect for $t_1=t_2=1$, Fig.~\ref{fig:onsitetuning}(a), where enhanced translational symmetry allows access to $L=24$ and maximizes intra-band hybridization beyond the anomalous boundary. For other parameter choices ($L=22$), finite-size limitations reduce the sharpness of the dip but preserve the intercept scaling.

For sufficiently large $h$, the ladder spacing is set by the sublattice gap $\sim 2h$, and minibands reorganize around local site occupation. When both $h$ and $V_{\rm int}$ become large, the Hamiltonian becomes approximately diagonal in the occupation basis, $H \approx h\sum_j (-1)^j n_j + V_{\rm int}\sum_j n_j n_{j+1}$, suppressing off-diagonal hybridization. In this regime $\langle r\rangle$ decreases due to emergent Fock-space fragmentation rather than incomplete miniband hybridization, and the anomalous line broadens into a wider region of reduced $\langle r\rangle$.

These results demonstrate that the anomalous statistics arise from a hierarchical MB miniband structure controlled by single-particle band geometry. Breaking chiral symmetry modifies the nature of the ladders but does not eliminate the mechanism, confirming its geometric rather than symmetry-protected origin.

\end{document}